\documentclass[12pt]{iopart}

\usepackage{iopams}

\expandafter\let\csname equation*\endcsname\relax

\expandafter\let\csname endequation*\endcsname\relax

\usepackage{amsmath}
\usepackage{graphicx}
\usepackage{dcolumn}
\usepackage{bm}
\usepackage{xcolor}

\def\txi{\tilde{\boldsymbol{\xi}}}
\def\tx{\tilde{\mathbf{x}}}
\def\mt{\mathcal{T}}

\begin{document}

\title[Entropy production in active Rouse polymers]{Entropy production in active Rouse polymers}

\author{Sandipan Dutta}

\address{Department of Physics, Birla Institute of Technology and Science, Pilani, Rajasthan, 333031, India}
\ead{sandipan.dutta@pilani.bits-pilani.ac.in}
\vspace{10pt}
\begin{indented}
\item[]July 2022
\end{indented}

\begin{abstract}
Active polymers are the archetype of nonequilibrium viscoelastic systems that constantly consume energy to produce motion. The activity of many biopolymers is essential to many life processes. The entropy production rate quantifies their non-equilibrium nature through the breaking of the time reversal symmetry. In this work we build an analytical model of active polymers as active Rouse polymers where the beads are active Ornstein–Uhlenbeck particles (AOUP) and calculate their entropy production. The interactions between the beads are decoupled through the normal mode analysis and the entropy production can be solved analytically. We obtain the contribution of each Rouse mode in the entropy production and the dependence of the entropy production on the polymer properties like length. We find that the entropy production is zero for a passive Rouse polymer in the presence of thermal bath as well as for an active Rouse polymer in the absence of thermal bath. For an active chain in the presence of a thermal bath the entropy production is non-zero. In this case we find that the local temporal entropy production dominates the non-local entropy production. 
\end{abstract}

%
%
%
%
%

\section{Introduction}

Active matter systems constantly dissipate energy, either stored internally or extracted from the environment, to perform mechanical work \cite{marchetti2013hydrodynamics,bechinger2016active, elgeti2015physics, vicsek2012collective}. Many biological systems display a variety of collective behaviour ranging from the flocking of birds, fishes and herds of animals, to swimming organisms like bacteria and algae \cite{ramaswamy2010mechanics}, to cytoskeleton in living cells \cite{kruse2004asters, juelicher2007active} and cells in tissues \cite{Dapeng2016}. Inspiration from the living world led researchers to develop artificial particles like Janus colloids \cite{howse2007self, jiang2010active, volpe2011microswimmers}, colloidal rollers \cite{bricard2015emergent} and water droplets \cite{izri2014self, thutupalli2011swarming} that provide more control over their motility. 

Active systems are inherently out of equilibrium which break detailed balance at microscopic scales and hence can not be described by the equilibrium Gibbsian statistical mechanics. Their collective behaviour is usually studied using hydrodynamic equations \cite{marchetti2013hydrodynamics} exploiting the symmetries and conservation of the system \cite{toner1998}. Many of the earlier works \cite{lau2009, speck2014, solon2015pressure, bialke2015, maggi2015multidimensional, trefz2016activity, fily2017equilibrium, brader2015}  described the thermodynamics of active matter by mapping them to passive systems. Stochastic thermodynamics \cite{seifert2012stochastic, jarzynski2013equalities} is often used to study the nonequilibrium systems where the thermodynamic quantities like heat, work and entropy are fluctuating quantities depending on the trajectories of the system. Several models have been used to study active systems like the active Brownian particle \cite{romanczuk2012active, ten2011brownian, sevilla2015smoluchowski}, run and tumble \cite{nash2010run, tailleur2008statistical, angelani2015run} and active Ornstein–Uhlenbeck particle (AOUP) model \cite{szamel2014self, maggi2015marini, martin2021statistical}. Out of these, the AOUP model provides a framework that is analytically tractable. The dynamics of the position $\{\mathbf{x}_i\}$ of a system of $N$ AOUP particles having an interaction potential $\Phi$ and a drag coefficient $\gamma$ is given by
\begin{equation}
\dot{\mathbf{x}}_i = -\frac{1}{\gamma}\boldsymbol{\nabla}_i\Phi + \boldsymbol{\xi}_i^A. 
\label{aoup}
\end{equation}
$\boldsymbol{\xi}_i^A$ is an Ornstein-Uhlenbeck process which satisfies $\langle \xi_{i\alpha}^A \rangle = 0$ and $\langle \xi_{i\alpha}^A\xi_{j\beta}^A\rangle = \delta_{ij}\delta_{\alpha\beta}\frac{D_a}{\tau}\exp(-\vert t\vert/\tau)$, where $\alpha$ and $\beta$ correspond to the spatial components. $D_a$ is the strength of the colored noise corresponding to active diffusion and $\tau$ the persistence time of the active motion.  

Of all the active systems, active polymers and filaments \cite{osmanovic2017dynamics, winkler2017active} form an interesting class of system where deformation plays an important role in the dynamics in addition to the propulsion and the thermal noise. Examples include enzymes catalysing chemical reactions \cite{muddana2010substrate, jee2018catalytic, jee2018enzyme}, the motion of the chromosomal loci \cite{weber2012nonthermal, javer2013short} caused by active microtubule \cite{brangwynne2008nonequilibrium} or actin \cite{weber2015random} and Janus chains \cite{yan2016reconfiguring}. Active polymeric fluids display a wide range of collective behavior like asters and vortices, spirals, and motile topological defects. Hydrodynamic theories and particle based models \cite{marchetti2013hydrodynamics, winkler2017active} have been used to study these emergent behavior. While the multi polymer systems have rich phase diagrams, it is worthwhile to understand the single polymer behaviour. Earlier works \cite{osmanovic2017dynamics,winkler2017active, sakaue2017active,  chaki2019enhanced, vandebroek2015dynamics, ghosh2014dynamics, samanta2016chain, eisenstecken2017internal, bianco2018globulelike,  winkler2020physics, gupta2021heat, santra2022activity} mostly focused on the polymer dynamics in athermal baths or the dynamics of active polymers in thermal bath. But few works exist on how their dynamics affect their thermodynamic properties like the entropy production. Entropy production is an important measure of how far the system is from equilibrium and characterizes the energy consumption of these active systems.  Here we study the entropy production of active Rouse polymers \cite{doi1988theory} in the presence and absence of thermal baths. The polymer consists of beads that are self-propelled AOUP particles connected by springs \cite{osmanovic2017dynamics}. The dynamics of the beads coupled by the spring interactions are decoupled using normal mode analysis. This enables us to analyse the contribution of each Rouse mode to the entropy production. Following \cite{caprini2019entropy} we find that the entropy production strongly depends on the environment of the polymer i.e. the presence or the absence of the bath. For an active polymer in a thermal bath we find that the local entropy production is much higher than the non-local entropy production rate.

The breaking of the time-reversal symmetry in active systems is quantified by the entropy production rate and is taken as the ratio of probabilities of the forward and the backward trajectories \cite{lebowitz1999gallavotti}. While there are several approaches to the calculation of the entropy production in active systems \cite{ganguly2013stochastic, nardini2017entropy, mandal2018mandal, dabelow2019irreversibility, shankar2018hidden, dabelow2021irreversible}, we follow the path-integral approach generalised to non-Markovian noise \cite{zamponi2005fluctuation, caprini2019entropy}. Caprini et. al. \cite{caprini2019entropy} have shown that this method gives the entropy production in AOUP particles that is consistent with other methods without any additional assumptions about the system. We use the formalism developed by Caprini et. al. \cite{caprini2019entropy} to study the entropy production in active polymers. Both Caprini et. al. and Dabelow et. al. \cite{dabelow2019irreversibility} obtained the entropy production for a single AOUP in an external potential. We generalise their results to Rouse polymers which are multi-particle systems interacting via harmonic potential. Our calculation of entropy production for the Rouse modes provides information about the collective dynamics of the active monomers that is missing in single particle systems. Each Rouse mode has distinct conformational dynamics. This enables us to analyze the role of geometry and deformation on the entropy production in active polymers.   

The paper is organised as follows. In Section \ref{sec:rouse} we describe the dynamics of an active Rouse chain as AOUP particles interacting through harmonic potentials. In Section \ref{sec:entropy} we define the entropy production in a trajectory within the path integral formalism of Ref \cite{caprini2019entropy, zamponi2005fluctuation}. The entropy production of a passive Rouse chain in a thermal bath and an active Rouse chain in the presence and absence of a thermal bath is calculated in Section \ref{sec:entropy polymers}. Finally we conclude in Section \ref{sec:conclusion}.

\section{The dynamics of a Rouse chain}
\label{sec:rouse}
First we look at the dynamics of a passive Rouse polymer in a thermal bath. Consider a Rouse chain of length $N$ connected by elastic springs of spring constant $\kappa$ in a thermal bath of temperature $T$. The dynamics of the internal beads $i = 2,...,N-1$ follow
\begin{equation}
\dot{\mathbf{x}}_{i}(t) = -\frac{\kappa}{\gamma}\biggl(2\mathbf{x}_i(t)-\mathbf{x}_{i-1}(t)-\mathbf{x}_{i+1}(t)\biggr) + \boldsymbol{\xi}^T_i(t),
\label{langevinthermal}
\end{equation}
where the thermal noise $\boldsymbol{\xi}^T$ satisfies $\langle\xi^T_{i,\alpha}(t)\rangle = 0$ and $\langle\xi^T_{i,\alpha}(t)\xi^T_{j,\beta}(s)\rangle = \frac{2 k_BT}{\gamma}\delta_{i,j}\delta_{\alpha,\beta}\delta(t-s)$. $\alpha$ and $\beta$ are the spatial component indices. For the end beads $i = 1$ and $N$ we have  
\begin{align}
&\dot{\mathbf{x}}_{1}(t) = -\frac{\kappa}{\gamma}(\mathbf{x}_{1}(t)-\mathbf{x}_{2}(t)) + \boldsymbol{\xi}^T_{1}(t), & \dot{\mathbf{x}}_{N}(t) = -\frac{\kappa}{\gamma}(\mathbf{x}_{N}(t)-\mathbf{x}_{N-1}(t)) + \boldsymbol{\xi}^T_{N}(t).
\end{align}
This equation can also be recast into Eq \eqref{langevinthermal} provided we define two additional hypothetical beads $\mathbf{x}_0$ and $\mathbf{x}_{N+1}$ with $\mathbf{x}_0=\mathbf{x}_1$ and $\mathbf{x}_{N+1}=\mathbf{x}_{N}$ \cite{doi1988theory}. The spring interactions have the form $\Phi(\{\mathbf{x}_i\}) = \sum_i\frac{\kappa}{2}(\mathbf{x}_i-\mathbf{x}_{i-1})^2$. Eq \eqref{langevinthermal} can be written in a matrix form as
\begin{equation}
\dot{\mathbf{X}} = \frac{1}{\gamma}\mathbf{F}(\mathbf{X})+\boldsymbol{\Xi}^T,
\label{langevin}
\end{equation}
with the $i$th element of $\mathbf{X}$ and $\boldsymbol{\Xi}$, $(\mathbf{X})_i=\mathbf{x}_i$ and $(\boldsymbol{\Xi})_i = \boldsymbol{\xi}_i$. $\mathbf{F}_i(\mathbf{X})=-\boldsymbol{\nabla}_i\Phi(\mathbf{X})$ and the noise covariance matrix of $\boldsymbol{\Xi}$ is
\begin{equation}
\boldsymbol{\nu}(t-s) = \frac{2k_BT}{\gamma}\mathbf{I}\delta(t-s),
\label{noisecovariance}
\end{equation}
where $\mathbf{I}$ is the identity matrix of size $3N$.
The coupled dynamics of the beads in Eq \eqref{langevinthermal} is solved using the Rouse modes by taking its continuous limit as detailed in \cite{doi1988theory}. The first term on the right hand side of the equation can be identified as the discrete Laplacian in the bead indices $i$. The continuous limit of $i$ in this term gives the Laplacian or the second derivative with respect to the bead index $i$, 
\begin{equation}
\dot{\mathbf{x}}_{i}(t) = \frac{\kappa}{\gamma}\frac{\partial^2}{\partial i^2}\mathbf{x}_i(t) + \boldsymbol{\xi}^T_i(t).
\label{langevinrouse}
\end{equation}
The boundary conditions comes from the definition of the hypothetical beads which in the continuous limit are: $\frac{\partial\mathbf{x}_i}{\partial i}\vert_{i = 0} = 0$ and $\frac{\partial\mathbf{x}_i}{\partial i}\vert_{i = N} = 0$. The continuous formalism would be valid in the long time-scale motion for which $\mathbf{x}_i$ varies slowly with $i$. More generally, a Langevin equation linear in $\mathbf{x}$ with localized interactions would yield the Rouse equation Eq \eqref{langevinrouse} at long times as shown in Ref \cite{doi1988theory}. The generalization to the active Rouse chain is straightforward. We model the beads of the polymer to be AOUP particles following the dynamics of Eq \eqref{aoup}. To study an active Rouse chain we add an athermal contribution $\boldsymbol{\xi}^A$ to the noise 
\begin{equation}
\dot{\mathbf{x}}_{i}(t) = \frac{\kappa}{\gamma}\frac{\partial^2}{\partial i^2}\mathbf{x}_i(t) + \boldsymbol{\xi}_i(t),
\label{langevinactive}
\end{equation}
where the noise $\boldsymbol{\xi} = \boldsymbol{\xi}^T +\boldsymbol{\xi}^A$. Within the AOUP model, the athermal component satisfies $\langle\xi^A_{i\alpha}(t)\rangle = 0$ and $\langle\xi^A_{i\alpha}(t)\xi^A_{j\beta}(s)\rangle = \delta_{ij}\delta_{\alpha\beta}\frac{D_a}{\tau}\exp(-\vert t-s\vert/\tau)$ as in Eq \eqref{aoup}.
The harmonic interactions among the beads can be decoupled using the Rouse modes defined by \cite{doi1988theory, osmanovic2017dynamics}
\begin{equation}
\tx_{p}(t) = \frac{1}{N}\int_0^Ndi\cos(p\pi i/N)\mathbf{x}_{i}(t).
\label{normalmodes}
\end{equation}
The particle positions can be written in terms of the Rouse modes as
\begin{equation}
\mathbf{x}_i(t) = \tx_0(t) + 2\sum_{p=1}^N\tx_p(t)\cos(p\pi i/N).
\label{inversenormalmodes}
\end{equation}
$\tx_0(t) = \frac{1}{N}\int_0^Ndi\mathbf{x}_i(t)$ represents the position of the center of mass of the polymer.These equations are the cos transform of the bead indices and the corresponding inverse transform. Using the transform (Eq \eqref{normalmodes}) of Eq \eqref{langevinrouse}, we obtain an uncoupled set of equations for the Rouse modes 
\begin{equation}
\dot{\tx}_{p}(t) = -\kappa_p\tx_p(t) + \tilde{\boldsymbol{\xi}}_p(t),
\label{normalmodelangavin}
\end{equation}
where $\kappa_p = \frac{3\pi^2 k_BT}{Nb^2\gamma}p^2$. $b$ is the monomer size obtained by taking the standard deviation of the distance between two beads. The derivation involves integration by parts where boundary terms vanish due to the conditions: $\frac{\partial\mathbf{x}_i}{\partial i}\vert_{i = 0} = 0$ and $\frac{\partial\mathbf{x}_i}{\partial i}\vert_{i = N} = 0$. The cos transform in Eq \eqref{normalmodes} emerges naturally by assuming that with an appropriate choice of $\phi_{pi}$ in general transform: $\tx_p(t) = \int_0^Ndi\phi_{pi}\mathbf{x}_i(t)$, yields a decoupled equation Eq \eqref{normalmodelangavin}. Using this transform in Eq \eqref{langevinactive} and integrating by parts with the boundary conditions at the end beads: $\frac{\partial\mathbf{x}_i}{\partial i}\vert_{i = 0} = 0$ and $\frac{\partial\mathbf{x}_i}{\partial i}\vert_{i = N} = 0$, we obtain $\frac{\partial^2\phi_{pi}}{\partial i^2} = -\Gamma_p\phi_{pi}$. $\Gamma_p = \gamma\kappa_p/\kappa$. Solution of this differential equation with the boundary conditions: $\frac{\partial\phi_{pi}}{\partial i} = 0$ at $i = 0$ and $N$, yield $\phi_{pi}=\frac{1}{N}\cos(p\pi i/N)$. The transformed
noises are obtained similar to Equation \eqref{normalmodes} 
\begin{equation}
\tilde{\xi}_{p,\alpha}^{A,T}(t) = \frac{1}{N}\int_0^Ndi\cos(p\pi i/N)\xi_{i,\alpha}^{A,T}(t) 
\end{equation}
From this we get 
\begin{align}
\langle\tilde{\xi}^{I}_{p,\alpha}(t)\rangle & = \frac{1}{N}\int_0^Ndi\cos(p\pi i/N)\langle\xi_{i,\alpha}^{I}(t)\rangle = 0,\nonumber\\
\langle\tilde{\xi}^{I}_{p,\alpha}(t)\tilde{\xi}^{I}_{q,\beta}(s)\rangle & = \frac{1}{N^2}\int_0^Ndi\int_0^Ndj\cos(p\pi i/N)\cos(q\pi j/N)\langle\xi^{I}_{i,\alpha}(t)\xi^{I}_{j,\beta}(s)\rangle\nonumber\\
& = \frac{1}{N^2}\int_0^Ndi\int_0^Ndj\cos(p\pi i/N)\cos(q\pi j/N)\delta_{ij}\delta_{\alpha\beta}f^I(t-s),\nonumber\\
\end{align}
where $I = A$ or $T$. Using the explicit forms of the noise correlations $f^T(t-s) = \frac{2k_BT}{\gamma}\delta(t-s)$ or $f^A(t-s) = \frac{D_a}{\tau}\exp(-\vert t-s\vert/\tau)$ and the orthogonality condition $\int_0^Ndi\cos(p\pi i/N)\cos(q\pi i/N) = \frac{N}{2}\delta_{pq} $, we can show that the transformed thermal noise satisfy
\begin{align}
&\langle\tilde{\xi}^T_{p\alpha}(t)\rangle = 0\text{,} & \langle\tilde{\xi}^T_{p\alpha}(t)\tilde{\xi}^T_{q\beta}(s)\rangle = \frac{2k_BT}{\gamma N}\delta_{pq}\delta_{\alpha\beta}\frac{1+\delta_{p0}}{2}\delta(t-s),
\label{thermalnormalmodenoise}
\end{align}
and the transformed athermal noise 
\begin{align}
&\langle\tilde{\xi}^A_{p\alpha}(t)\rangle = 0\text{,}  &\langle\tilde{\xi}^A_{p\alpha}(t)\tilde{\xi}^A_{q\beta}(s)\rangle = 2\delta_{pq}\delta_{\alpha\beta}\frac{1+\delta_{p0}}{2N}\frac{D_a}{\tau}\exp(-\vert t-s\vert/\tau).
\label{athermalnormalmodenoise}
\end{align}
The matrix form of the Langevin equations for the Rouse modes in Equation \eqref{normalmodelangavin} reads
similar to Equation \eqref{langevin}
\begin{equation}
\dot{\widetilde{\mathbf{X}}} = \frac{1}{\gamma}\widetilde{\mathbf{F}}(\widetilde{\mathbf{X}}) + \widetilde{\boldsymbol{\Xi}}.
\end{equation} 
The $p$th element of $(\widetilde{\mathbf{X}})_p = \widetilde{\mathbf{x}}_p$ and $(\widetilde{\boldsymbol{\Xi}})_p = \widetilde{\boldsymbol{\xi}}_p$.

\section{Entropy production}
\label{sec:entropy}
 We denote the positions of the beads of an $N$ bead polymer from Equation \eqref{langevin} by $\mathbf{X}$. Caprini et. al. \cite{caprini2019entropy} have obtained the probability of a trajectory of the polymer from the initial point $\mathbf{X}_0$ to $\mathbf{X}$ for time $\mt$ in a path integral form 
\begin{equation}
\mathcal{P}[\mathbf{X}\vert\mathbf{X}_0] \propto \exp\left[-\frac{1}{2}\int_0^{\mt} ds \int_0^{\mt} dt[\dot{\mathbf{X}}-\mathbf{F}(\mathbf{X})](s)\mathbf{T}^{-1}(s-t)[\dot{\mathbf{X}}-\mathbf{F}(\mathbf{X})](t)\right],
\label{trajectoryprobability}
\end{equation}
where $\mathbf{T}$ is the inverse of the covariance matrix in Equation \eqref{noisecovariance}
\begin{equation}
\int dt^{\prime}\mathbf{T}^{-1}(t-t^{\prime})\boldsymbol{\nu}(t^{\prime}-s) = \mathbf{I}\delta(t-s).
\label{covarianceinverse}
\end{equation}
$\mathbf{T}$ is interpreted as an effective temperature of the system that contains the information of the non-Markovian nature of the bath \cite{chaki2019effects, zamponi2005fluctuation}. The time reversed trajectory $\overleftarrow{\mathbf{X}}$ is obtained by $\overleftarrow{\mathbf{X}}(t)= \mathbf{X}({\mt}-t)$. Since $\mathbf{X}$ and $\mathbf{T}$ are even under time reversal ($\overleftarrow{\mathbf{X}}(t)= \mathbf{X}(t)$), the probability of time reversed trajectory is
\begin{equation}
\mathcal{P}[\overleftarrow{\mathbf{X}}\vert\overleftarrow{\mathbf{X}}_0] \propto \exp\left[-\frac{1}{2}\int_0^{\mt} ds \int_0^{\mt} dt[-\dot{\mathbf{X}}-\mathbf{F}(\mathbf{X})](s)\mathbf{T}^{-1}(s-t)[-\dot{\mathbf{X}}-\mathbf{F}(\mathbf{X})](t)\right].
\label{reversetrajectoryprobability}
\end{equation}
The entropy production of the medium is given by the ratio of the log of the probability of the forward and backward trajectory \cite{zamponi2005fluctuation, seifert2012stochastic} 
\begin{equation}
\Sigma = \log\frac{\mathcal{P}[\mathbf{X}\vert\mathbf{X}_0]}{\mathcal{P}[\overleftarrow{\mathbf{X}}\vert\overleftarrow{\mathbf{X}}_0]}.
\end{equation}

Using the explicit forms of the probabilities in Equation \eqref{reversetrajectoryprobability} and Equation \eqref{trajectoryprobability}, we obtain the entropy production of the medium as \cite{caprini2019entropy}
\begin{equation}
\Sigma = \int_0^{\mt} dt\sigma(t),
\label{entropyproduction}
\end{equation}
where the time dependent entropy production rate is
\begin{equation}
\sigma(t) = \int_0^{\mt} ds\left[\dot{\mathbf{X}}(t)\mathbf{T}^{-1}(t-s)\mathbf{F}(\mathbf{X}(s))+\mathbf{F}(\mathbf{X}(t))\mathbf{T}^{-1}(t-s)\dot{\mathbf{X}}(s)\right].
\label{entropyrate}
\end{equation}
 From now on we will use the term entropy production for the entropy production of the medium.

\section{Entropy production in polymers}
\label{sec:entropy polymers}

\subsection{Passive Rouse chain in thermal bath}
Now we apply the above formalism to obtain the entropy production of active/passive Rouse polymers in presence/absence of a thermal bath. First we consider the case of a Rouse chain with passive beads in a thermal bath. From Equation \eqref{noisecovariance} and Equation \eqref{covarianceinverse} we obtain
\begin{equation}
\mathbf{T}^{-1}(t-s) = \frac{\gamma}{2k_BT}\mathbf{I}\delta(t-s).
\end{equation}
Using this in the entropy production rate from Equation \eqref{entropyrate} we obtain
\begin{align}
\sigma(t) & = \frac{\gamma}{ k_BT}\dot{\mathbf{X}}.\frac{1}{\gamma}\mathbf{F}(\mathbf{X}) =-\frac{1}{ k_BT}\dot{\mathbf{X}}.\nabla \Phi(\mathbf{X}) = -\frac{1}{ k_BT}\frac{d}{dt}\Phi(\mathbf{X}).
\label{entropypassive}
\end{align}
Since this is a time derivative and its integral is a boundary term, the average entropy production in Equation \eqref{entropyproduction}  vanishes. In an uncorrelated thermal bath, there is no production of entropy. 

The Rouse modes of the polymers represent the collective behavior of the polymer. We investigate the entropy production in each of these modes from Equation \eqref{langevinthermal} and Equation \eqref{entropypassive}. Equation \eqref{entropypassive} becomes
\begin{align}
\sigma(t) & \approx \frac{\kappa}{ k_BT}\int_{0}^Ndi\dot{\mathbf{x}}_i.\frac{\partial^2}{\partial i^2}\mathbf{x}_i \nonumber\\
& = -\frac{2}{k_BT}\int_0^N di[\dot{\tx}_0+2\sum_p\dot{\tx}_p\cos(p\pi i/N)].[k_q\sum_q\tx_q\cos(q\pi i/N)] \nonumber\\
& = -\frac{2N}{k_BT}\sum_{p=1}^Nk_p\dot{\tx}_p.\tx_p = -\frac{N}{ k_BT}\frac{d}{dt} \biggl(\sum_{p=1}^Nk_p\tx^2_p\biggr),
\label{entropythermal}
\end{align}
where $k_p^2 = \frac{\kappa\pi^2p^2}{\gamma N^2} = \frac{3\pi^2k_BTp^2}{b^2N^2}$. The average entropy production rate for mode $p$ is obtained from Equation \eqref{msd} by setting $D_a = 0$  
\begin{align}
\langle\sigma_p(t) \rangle & = -\frac{N}{ k_BT}k_p\frac{d}{dt}\langle\tx^2_p\rangle \nonumber\\
&= 0,    
\end{align}
where we have used $\langle \tx^2_p\rangle = \frac{k_BT}{k_p}$. Thus the total entropy production is again zero.

\subsection{Active Rouse chain in absence of thermal bath}
\begin{figure}
  \includegraphics[width=\linewidth]{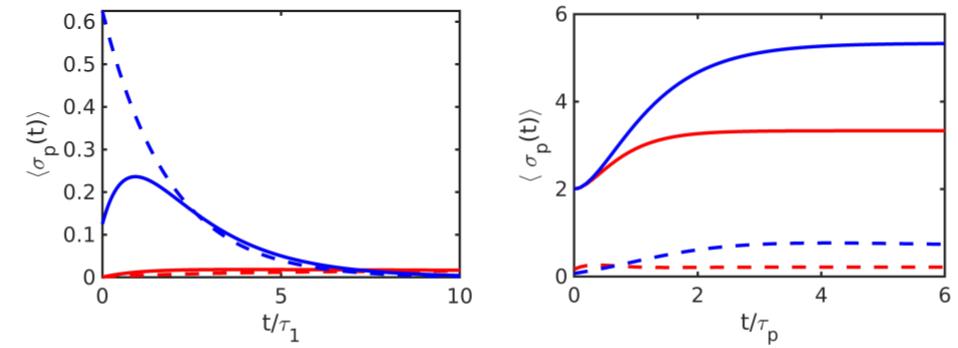}
  \caption{The average entropy production rate in Equation\ref{eq:nothermal} in the absence of thermal bath for a polymer with $N = 10$ monomers for mode $1$ (red) and $5$ (blue) and $\tau = 1$ (solid) and $\tau = 5$ (dashed). The times are scaled with respect to a monomer time $\tau_1$. The second plot shows the local (solid) and non-local (dashed) average entropy production in Equation\ref{eq:thermal} in the presence of a thermal bath for $D_a = 1, \tau = 0.5$ (red) and $D_a = 10, \tau = 5$ (blue).}
  \label{fig:active}
\end{figure}

Now we consider the case of active Rouse polymers in the absence of a thermal bath. This is the case of active systems where the thermal diffusion is negligible compared to the motion due to self-propulsion. The thermal noise $\txi^T$ is set to zero in Equation \eqref{langevinactive}. $\txi^A$ represents the self-propulsion term for AOUP dynamics, which is Gaussian and is exponentially correlated in time. The noise covariance matrix is
\begin{equation}
\boldsymbol{\nu}(t-s) = \frac{D_a}{\tau}\exp(-\vert t-s\vert/\tau)\mathbf{I},
\end{equation}
which gives in Ref \cite{caprini2019entropy}
\begin{equation}
\mathbf{T}^{-1}(t-s) = \frac{\delta(t-s)}{2D_a}\left(1-\tau^2\frac{d^2}{dt^2}\right)\mathbf{I}.
\end{equation}
The entropy production rate can be calculated from Equation \eqref{entropyrate}
\begin{align}
\sigma(t) 
& = -\frac{1}{2D_a\gamma}\sum_{i=1}^N\left[\dot{\mathbf{x}}_i(t).\int ds \delta(t-s) \left(1-\tau^2\frac{d^2}{ds^2}\right)\nabla_i\Phi(s)+\nabla_i\Phi(t). \int ds\delta(t-s) \left(1-\tau^2\frac{d^2}{ds^2}\right)\dot{\mathbf{x}}_i(s)\right]\nonumber\\
& \approx \frac{\kappa}{2D_a\gamma}\int_0^Ndi\left[\dot{\mathbf{x}}_i(t).\left(1-\tau^2\frac{d^2}{dt^2}\right)\frac{\partial^2}{\partial i^2}\mathbf{x}_i(t)+\frac{\partial^2}{\partial i^2}\mathbf{x}_i(t). \left(1-\tau^2\frac{d^2}{dt^2}\right)\dot{\mathbf{x}}_i(t)\right]\nonumber\\
& \approx -\frac{N}{D_a}\sum_{p=1}^Nk_p\left[\dot{\tx}_p(t).\left(1-\tau^2\frac{d^2}{dt^2}\right)\tx_p(t)+\tx_p(t). \left(1-\tau^2\frac{d^2}{dt^2}\right)\dot{\tx}_p(t)\right]\nonumber\\
& = -\frac{N}{D_a}\frac{d}{dt}\biggl(\sum_{p=1}^Nk_p[\tx_p^2(t)-\tau^2\tx_p(t).\ddot{\tx}_p(t)]\biggr).
\end{align}
Since this is an exact time derivative the total entropy production is again zero. The average entropy production in mode $p$ is 
\begin{align}
\langle\sigma_p(t)\rangle & =-\frac{Nk_p}{D_a}\frac{d}{dt}[\langle\tx_p^2(t)\rangle-\tau^2\langle\tx_p(t).\ddot{\tx}_p(t)\rangle] \\
& =-\frac{Nk_p}{D_a}\frac{d}{dt}[(1-\frac{\tau^2}{\tau_p^2})\langle\tx_p^2(t)\rangle+ \frac{\tau^2}{\tau_p}\langle\txi_p(t).\tx_p(t)\rangle-\tau^2\langle\dot{\txi}_p(t).\tx_p(t)\rangle\rangle] \\
& = -\frac{Nk_p}{D_a}\frac{d}{dt}\biggl[\frac{D_a(\tau_p-\tau)}{N} \biggl(1-\exp(-2t/\tau_p)\biggr)+\frac{2D_a\tau}{N}\biggl(\exp(-(\frac{1}{\tau_p}+\frac{1}{\tau})t)-\exp(-2t/\tau_p)\biggr) \nonumber\\& + \frac{2D_a\tau^2}{N(\tau+\tau_p)}\biggl(1-\exp(-(\frac{1}{\tau_p}+\frac{1}{\tau})t)\biggr)\biggr]\nonumber\\
& = 2k_p\biggl[\exp(-(\frac{1}{\tau}+\frac{1}{\tau_p})t)-(1+\frac{\tau}{\tau_p})\exp(-\frac{2}{\tau_p}t)\biggr],
\label{eq:nothermal}
\end{align}
where we have used Equation \eqref{xddotexpansion} in the derivation. The average entropy production in each mode does not depend on $D_a$ in absence of thermal noise. The first plot of Figure \ref{fig:active} shows the average entropy production is higher for larger persistence times $\tau$  and higher order modes $p$.   

\begin{figure}
  \includegraphics[width=\linewidth]{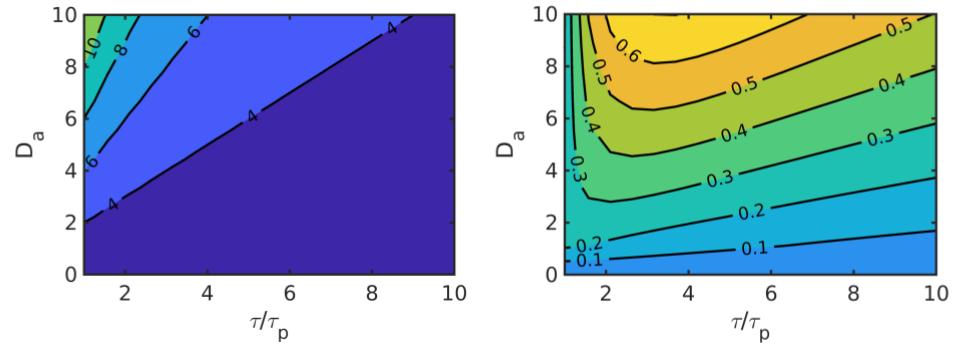}
  \caption{The phase space of the steady state local (solid) and non-local (dashed) entropy production rate vs $D_a$ and $\tau$. }
  \label{fig:entropy}
\end{figure}

\subsection{Active Rouse chain in thermal bath}
Most active systems have self-propelled particles swimming in a solvent. The solvent is assumed to be in equilibrium and hence is modeled as a thermal bath which is unaffected by the swimming polymers. In this case the noise in Equation \eqref{langevinactive} has $\txi^T$ due to the thermal bath and $\txi^A$ originating from the self-propulsion. The thermal noise is memory-less $\langle\txi^T(t)\txi^T(s)\rangle=\frac{2k_BT}{\gamma}\delta_{\alpha\beta}\delta(t-s)$, while the athermal or active noise is correlated with a persistence time $\tau$, $\langle\txi^A(t)\txi^A(s)\rangle=\frac{D_a}{\tau}\delta_{\alpha\beta}\exp(-\vert t-s\vert/\tau)$, where the indices $\alpha, \beta$ are the $x, y$ and $z$ coordinates. The polymer is effectively in contact with two reservoirs, thermal and colored noise bath. For this combined bath from Ref \cite{caprini2019entropy}
\begin{equation}
\mathbf{T}^{-1}(t) = \frac{\gamma}{2k_BT}\mathbf{I}+K^{-1}(t)\mathbf{I},
\end{equation}
where $K^{-1}(t) = -\frac{D_a\gamma^2}{4k_B^2T^2\tau}\frac{1}{\sqrt{1+D_a\gamma/k_BT}}\exp(-\frac{\vert t\vert}{\tau}\sqrt{1+D_a\gamma/k_BT})$.

The entropy production rate reads
\begin{align}
\sigma(t) & = -\frac{1}{\gamma}\sum_{i=1}^N\dot{\mathbf{x}}_i(t).\int ds\biggl(\frac{\gamma}{2k_BT}\delta(t-s)+K^{-1}(t-s)\biggr)\nabla_i \Phi(s)-\frac{1}{\gamma}\nabla_i\Phi(t).\int ds\biggl(\frac{\gamma}{2k_BT}\nonumber\\&\delta(t-s)+K^{-1}(t-s)\biggr)\dot{\mathbf{x}}_i(s)\nonumber\\
& = \frac{\kappa}{k_BT\gamma}\int_0^Ndi\dot{\mathbf{x}}_i(t).\frac{\partial^2}{\partial i^2}\mathbf{x}_i(t)+\frac{\kappa}{\gamma}\int_0^Ndi\biggl[\dot{\mathbf{x}}_i(t)\int dsK^{-1}(t-s)\frac{\partial^2}{\partial i^2}\mathbf{x}_i(s)+\frac{\partial^2}{\partial i^2}\mathbf{x}_i(t) \nonumber\\&\int dsK^{-1}(t-s)\dot{\mathbf{x}}_i(s)\biggr] \nonumber\\
& = -2N\sum_{p=1}^Nk_p\biggl[\frac{1}{k_BT}\dot{{\tx}}_p(t).{\tx}_p(t)+\int dsK^{-1}(t-s)[\dot{\tx}_p(t).\tx_p(s)+\tx_p(t).\dot{\tx}_p(s)]\biggr].
\end{align}
The average entropy production rate for mode $p$ is
\begin{align}
\langle\sigma_p(t)\rangle & = -2Nk_p\biggl[\frac{1}{k_BT}\langle\dot{{\tx}}_p(t).{\tx}_p(t)\rangle+\int dsK^{-1}(t-s)[\langle\dot{\tx}_p(t).\tx_p(s)\rangle+\langle\tx_p(s).\dot{\tx}_p(t)\rangle]\biggr]
\label{eq:thermal}
\end{align}
The first term of the entropy rate is local in time while the second term is non-local. The second plot of Figure \ref{fig:active} shows that in the presence of the thermal bath the average entropy production rate reaches a maximum value, unlike in the case of the absence of the thermal bath where it decays to zero. The local entropy production dominates the non-local entropy production. The phase space of the maximum entropy rate for the local (first plot in Figure \ref{fig:entropy}) and non-local terms (second plot in Figure \ref{fig:entropy}) is shown for parameters $\tau$ and $D_a$. At short times $K(t)$ approaches a Dirac delta function (local) and the non-local interactions become large.

\section{Conclusions}
\label{sec:conclusion}
We have studied the entropy production of an active polymer in the presence and the absence of an equilibrium bath. Active polymeric systems are ubiquitous in biological systems in particular inside the cell where ATP is continuously consumed to translate or deform the biopolymers \cite{vignali2000atp, bruinsma2014chromatin}. Using the N-particle generalization of the path integral method to non-Markovian noises through effective temperature \cite{caprini2019entropy, zamponi2005fluctuation}, we have studied their entropy production. The interacting systems are not easily solvable due to their coupled dynamics, however the Rouse dynamics can be decoupled through the normal mode analysis. This along with the AOUP model makes the thermodynamics of the active Rouse chains analytically tractable. We calculated the entropy production for the Rouse modes that gives us an understanding of how the collective behavior of the active monomers and the geometry and the deformation of the polymers play a role in their entropy production.  While the entropy production is zero for a passive Rouse polymer in an equilibrium bath, it is non-zero for active polymers. For active polymers in a thermal bath we find that the local temporal interactions dominate the non-local interactions. Our work can be easily generalized to study the role of shear flow or external stresses on the active bioploymers like enzymes or chromatin.     

\ack 
The author acknowledges financial support by DST-SERB, India through the Startup Research Grant SRG/2022/000598 and MATRICS Grant MTR/2022/000281. The author also acknowledges financial support from the Additional Competitive Research Grant (PLN/AD/2022-23/3) from Birla Institute of Technology and Science, Pilani. The author also thanks DST-FIST for the computational resources provided to the Department of Physics, BITS Pilani.

\appendix 
\section{Calculation of correlations between normal modes}
\label{msd}
From Equation \eqref{normalmodelangavin} we get 
\begin{equation}
\tx_p(t) = \tx_p(0)\exp(-k_pt) + \int_0^tdt^{\prime}\exp(-k_p(t-t^{\prime}))\txi_p(t^{\prime}).
\label{xexpansion}
\end{equation}
From this with the assumption $t\ge s$ and the definition $\tau_p = 1/k_p$ we obtain
\begin{align}
\langle\tx_p(t).\tx_q(s)\rangle & = \langle\tx_p(0).\tx_q(0)\rangle\exp(-k_p(t+s))+\exp(-k_p(t+s))\int_0^tdt^{\prime}\int_0^sdt^{\prime\prime}\nonumber\\&\exp(k_p(t^{\prime}+t^{\prime\prime}))\left(\langle\txi_p^T(t^{\prime}).\txi_q^T(t^{\prime\prime})\rangle+\langle\txi_p^A(t^{\prime}).\txi_q^A(t^{\prime\prime})\rangle\right)\nonumber\\
& = \delta_{pq}\frac{k_BT}{Nk_p\gamma}\exp(-(t+s)/\tau_p)+\delta_{pq}\frac{k_BT}{Nk_p\gamma}(\exp(-\vert t-s\vert)/\tau_p)-\exp(-(t+s)/\tau_p))+\nonumber\\
& \frac{D_a}{\tau N}\delta_{pq}\int_0^tdt^{\prime}\int_0^sdt^{\prime\prime}dt^{\prime\prime}\exp((t^{\prime}+t^{\prime\prime})/\tau_p)\exp(-\frac{1}{\tau}\vert t^{\prime}-t^{\prime\prime}\vert)\nonumber\\
& = \delta_{pq}\frac{k_BT}{Nk_p\gamma}\exp(-\vert t-s\vert/\tau_p)+\frac{D_a}{\tau N}\delta_{pq}\exp(-(t+s)/\tau_p)\int_0^tdt^{\prime}\biggl[\int_0^{t^{\prime}}dt^{\prime\prime}\exp((t^{\prime}+t^{\prime\prime})/\tau_p)\nonumber\\&\exp(-\frac{1}{\tau}( t^{\prime}-t^{\prime\prime}))+\int_{t^{\prime}}^sdt^{\prime\prime}\exp((t^{\prime}+t^{\prime\prime})/\tau_p)\exp(-\frac{1}{\tau}( t^{\prime\prime}-t^{\prime}))\biggr]\nonumber\\
& = \delta_{pq}\frac{k_BT}{N k_p\gamma}\exp(-\frac{1}{\tau_p}\vert t-s\vert) +\frac{D_a}{\tau N(\frac{1}{\tau_p^2}-\frac{1}{\tau^2})}\delta_{pq}\biggl[-\frac{\tau_p}{\tau}\biggl(\exp(-\frac{1}{\tau_p}\vert t-s\vert)-\exp(-\frac{1}{\tau_p}(t+s))\biggr)\nonumber\\
&+\exp(-\frac{1}{\tau_p}(t+s))-\exp(-\frac{t}{\tau_p}-\frac{s}{\tau})-\exp(-\frac{t}{\tau}-\frac{s}{\tau_p})+\exp(-\frac{1}{\tau}\vert t-s\vert)\biggr].
\label{msd}
\end{align}

From Equation \eqref{normalmodelangavin}, Equation \eqref{xexpansion} and Equation \eqref{msd} we get
\begin{align}
\langle\dot{\tx}_p(t).\tx_q(s)\rangle & = -\frac{1}{\tau_p}\langle\tx_p(t).\tx_q(s)\rangle+\langle\txi_p(t).\tx_q(s)\rangle \nonumber\\
&=-\frac{1}{\tau_p}\langle\tx_p(t).\tx_q(s)\rangle+\int_0^sdt^{\prime}\exp(-(s-t^{\prime})/\tau_p)\left(\langle\txi_p^T(t).\txi_q^T(t^{\prime})\rangle+\langle\txi_p^A(t).\txi_q^A(t^{\prime})\rangle\right)\nonumber\\
&=-\frac{1}{\tau_p}\langle\tx_p(t).\tx_q(s)\rangle+\Theta(s\ge t)\delta_{pq}\biggl[\frac{k_BT}{N\gamma}\exp(-\frac{1}{\tau_p}(s-t))+D_a\frac{N\tau_p}{\tau}\biggl(1-\exp(-\frac{1}{\tau_p}(s-t))\biggr)\biggr] \nonumber\\& + \delta_{pq} \frac{D_a}{N(1+\tau/\tau_p)}\biggl(\exp(-\frac{1}{\tau_p}\vert s-t\vert)-\exp(-\frac{s}{\tau_p}-\frac{t}{\tau})\biggr).
\label{xdotexpansion}
\end{align}

For an active polymer in the absence of thermal noise using Equation \eqref{normalmodelangavin} and Equation \eqref{xdotexpansion} we have
\begin{align}
\langle\ddot{\tx}_p(t).\tx_p(t)\rangle & = \frac{1}{\tau_p^2}\langle\tx_p^2(t)\rangle - \frac{1}{\tau_p}\langle\txi_p(t).\tx_p(t)\rangle+\langle\dot{\txi}_p(t).\tx_p(t)\rangle\nonumber\\
& = \frac{1}{\tau_p^2}\langle\tx_p^2(t)\rangle-\frac{2D_a}{N(\tau+\tau_p)}\biggl(1-\exp(-(\frac{1}{\tau_p}+\frac{1}{\tau})t)\biggr)
\label{xddotexpansion}
\end{align}

\section*{References}
\bibliography{apssamp}
\bibliographystyle{unsrt}

\end{document}